\documentclass[12pt,A4,fleqn]{article}    % Specifies the document style.
\usepackage{epsfig}
\usepackage{amssymb,amsmath,amsthm}
\usepackage[title]{appendix}
\usepackage{tcolorbox}
\usepackage{caption}
\usepackage{bm}
\usepackage{rotating}
\usepackage{multirow}
\usepackage{hyperref}
\usepackage{float}
\usepackage{natbib}
\usepackage{algorithm}
\usepackage{algpseudocode}
\usepackage{diagbox}
\usepackage{parskip}
\usepackage{threeparttable}

\numberwithin{table}{section}
\numberwithin{equation}{section}
\numberwithin{figure}{section}
\usepackage{sectsty}
\subsectionfont{\small\selectfont}
\sectionfont{\normalsize \selectfont}
\subsubsectionfont{\footnotesize \selectfont}

\newtheorem{prop}{Proposition}[section]

\newcommand{\B}{\boldsymbol}

\newcommand\MyHead[2]{%
	\multicolumn{1}{l}{\parbox{#1}{\centering #2}}
}
\DeclareMathOperator{\E}{\mbox{E}}
\DeclareMathOperator{\Var}{\mbox{Var}}
\DeclareMathOperator{\sd}{\mbox{sd}}

\DeclareMathOperator*{\argmin}{argmin }

\title{How much is optimal reinsurance degraded by error?}
\author{Yinzhi Wang and Erik B\o lviken\\
Department of Mathematics\\
University of Oslo} 
\date{\today}

\begin{document}
\maketitle

\noindent
\small 
\newpage
\noindent
{\bf Abstract}
\\\\
The literature on optimal reinsurance does not  deal
with how much the effectiveness of such solutions are degraded
by errors in parameters and models. The issue
is investigated through both asymptotics and numerical studies.
It is shown that the rate of degradation is often $O(1/n)$
as the the sample size $n$ of historical observations becomes infinite. 
Criteria based on Value at Risk are exceptions that may
achieve only $O(1/\sqrt{n})$. These theoretical results are supported 
by 
numerical studies. A Bayesian perspective on how to integrate
risk caused by  parameter error 
is offered as well.
\\\\
{\bf Key words and phases}
\\
Asymptotics, Bayesian,  Conditional Value at Risk, frequentist, 
risk over expected surplus, Value at Risk. 
\newpage
\section{Introduction}
\noindent
Reinsurance is extensively used by insurance companies to reduce net
risk exposure and lower the reserve. This yields savings in capital cost  
which must be balanced against reinsurance expenses, 
and this creates an issue
of optimality as to what is the best trade-off. The problem was first 
attacked by \cite{borch1960} who showed that  stop-loss reinsurance
minimizes the variance of the expected loss for a given level of reinsurance,
and \cite{arrow1963} arrived at the same type of contract by 
maximizing the expected
utility of a risk-aversive insurer's terminal wealth. 
Both Borch and Arrow assumed
reinsurance premium to be proportional to the expected reinsurance pay-out,
the so-called expected premium principle.
The lack of realism here was realized by Borch himself, and it is not
surprising that there have in 
recent decades been a considerable upsurge of contributions based on other ways of pricing reinsurance, for example \cite{young1999, kaluszka2001, chi2013optimal} and \cite{cong2016}. In practice such premia depend strongly on the state of the market and may be highly fluctuating from one year to another. An insurance company  would from offers it has received from reinsurers
know something about the pricing schemes it faces, but such information is not publicly available, and academic work must therefore use so-called premium principles as proxies for market prices, as we do in this paper. A long list of them has been compiled in \cite{young2006premium}.

Then there is the question of how the trade-off between net reserve
and reinsurance cost should be put in mathematical form.
Many possibilities have found their way into actuarial literature, for
example \cite{kaluszka2004,cai2008optimal,balbas2009optimal} 
and \cite{cheung2014optimal} have minimized retained loss 
under some risk function;
whereas \cite{GAJEK2004227} and \cite{GUERRA2008529} 
maximize expected utility of wealth under different utility functions.
Much of the present  paper is concerned with  Value at Risk (VaR)
and Conditional Value at Risk (CVaR) against the insurer's expected profit.
CVaR is a coherent risk measure
and is much more in vogue
by theorists 
than VaR which does not satisfy the sub-additivity property; consult
\cite{Artzner1999}. Yet Value at Risk is arguably 
the more 
important from an industry point view since 
it is under 
current regulatory schemes directly linked to the cost of capital. 
Single layer contracts 
(i.e. excess of loss with an upper limit) may under 
under either risk measure be optimal
for single risks or at least close to that. Precise results of this nature
was established by
\cite{cheung2014optimal} under 
the expected
premium principle whereas \cite{chi2017optimal} 
under much more general conditions arrive at
multi-layer contracts, but those often reduce, at least approximately,
to single-layer ones
under  certain plausible constraints on the
reinsurance pricing functions introduced
in \cite{erik2019}. Some of these results
 are reviewed in Section \ref{sec:basics} as motivation for
the subsequent error study based on single-layer contracts. 

What is not known at all is to what extent
the optimality is upheld when there are 
errors in models and parameters. 
Optimal contracts are derived under estimated parameters or under
a postulated claim size distribution
that can't really be justified, and the solutions
are no longer optimal under the true
parameters or distribution. The question is how far
from the optimum we have now moved. 
Are criteria so sensitive that the solutions become very bad
or do they on the contrary remain close to the optimum?
How much historical data are needed to fit
parameters and distributions? The issue is a question of degradation
with contracts derived under estimated models evaluated 
under the true one so that it is possible to investigate how much worse
they have become.
Asymptotic studies as the number of historical observations  $n$ 
becomes infinite are carried out in Section \ref{sec3}. It will turn out that
the degradation rate is often $O(1/n)$ rather than the more usual 
$O(1/\sqrt{n})$, but in important special cases 
only the latter can be achieved.
The
coefficient of the leading error term is identified, and it is possible
to use it operatively  for numerical approximation, but 
when risk has to be computed by Monte Carlo in the first place,
it is often just as easy to implement a bootstrap (which amounts to nested 
simulations). The numerical study presented in Section \ref{sec:numercial1} is 
makes use of this tool. 
A Bayesian perspective is offered in Section \ref{sec:bayes}
and compared to the frequentist one numerically. 

\section{Preliminaries}\label{sec:basics}
\subsection{Notation and formulation}\label{subsec:notations}
Let $X$ be the total claim losses of a single portfolio
of non-life insurance policies over a certain period of time 
(often one year) and let $I=I(X)$ be the loss ceded to a reinsurer.
Natural restrictions on $I(x)$ are
\begin{equation}
0\leq I(x)\leq x
\quad \mbox{and} \quad
0\leq I(x_2)-I(x_1)\leq x_2-x_1
\quad\mbox{if} \quad x_1\leq x_2,
\label{e21}
\end{equation}
where the first condition is obvious since the reinsurer will never 
pay out more than the original claim. The second condition, 
known as the slow growth property is there to avoid moral hazard; consult \cite{chi2011optimal}. It is equivalent to a derivative $dI(x)/dx$
between
$0$ and $1$ where it exists, and it is crucial for the optimum results 
cited in Section \ref{subsec:2.5}.

The  retained risk of the insurer is 
\begin{equation}
R_I(X)=X-I(X)
\label{n1}
\end{equation}
with the subscript $I$ denoting the quantity to be optimized over.
Associated with $R_I(X)$ there is a risk measure, for example
Value at Risk (VaR) or Conditional Value at Risk (CVaR). 
Although it will in Section \ref{sec3} be necessary to highlight 
that these quantities depend on an underlying parameter vector
$\B\theta$ of the distribution function
$F(x;{\B\theta})$ of $X$, 
we can do without that for now. 
Their mathematical definitions at level $\epsilon$
are then
\begin{equation}
\mbox{VaR}_{R_I}=\inf\{x|1-F_{R_I}(x)\leq \epsilon\}
\quad \mbox{and} \quad
\mbox{CVaR}_{R_I}=E\{R_I(X)|R_I(X)\geq\mbox{VaR}_{R_I}\}
\label{e22}
\end{equation}
with $F_{R_I}(x)$ 
the distribution function of $R_I(X)$. Generic symbol
for risk measures in this paper is $\rho_{R_I}$.

The optimum problem considered in most of this 
paper is the trade-off 
between 
a risk measure and the expected surplus of the insurer 
for which a mathematical expression 
under a given
reinsurance treaty must be developed. If
$\pi$ is the premium collected from clients
and $\pi_I$ the reinsurance premium under $I(X)$, the 
economic summary of the operations is
\begin{equation}
A_I=\pi-X+I(X)-\pi_I-\beta\rho_{R_I},
\label{e27}
\end{equation}
where the last term on the right takes into account the cost
of  holding solvency capital through the cost of capital rate
$\beta\geq 0$. Note that this formulation
attaches cost to the entire net solvency capital
$R_I(X)$, not only to the 
part above the average as in \cite{chi2017optimal}.
Our choice  seems to us industrially plausible.
Let $G_I=E(A_I)$ be the expected surplus of the reinsurer. 
Taking expectations in~(\ref{e27}) yields 
\begin{equation}
G_I=\{\pi-E(X)\}-\{\pi_I-E\{I(X)\}\}-\beta \rho_{R_I},
\label{e28}
\end{equation}
which subtracts the expected surplus of the reinsurer and the capital cost from
the expected surplus of the insurer when no reinsurance has been bought.

\subsection{Premia}\label{subsec:prem}
In their simplest form premia are based on fixed loadings
$\gamma$ and $\gamma_r$ (both positive)  so that
\begin{equation}
\pi=(1+\gamma)E(X)
\quad\mbox{and}\quad
\pi_I=(1+\gamma_r)E\{I(X)\}
\label{e24}
\end{equation}
with 
$\gamma_r>\gamma$ in practice. 
The reinsurance part is inadequate since 
prices in that  market is likely to increase with risk
beyond a fixed 
coefficient $\gamma_r$. A more general formulation, used for example 
in \cite{chi2017optimal,erik2019}, is to introduce a market factor $M(Z)$ so that
\begin{equation}
\pi_I=E\{I(X)M(Z)\}.
\label{e25}
\end{equation}
Here $Z$ is a positive random variable correlated with $X$. The dependence
between $X$ and $Z$ is typically captured by a bivariate copula. 
Possible formulations of 
the market factor can be found in \cite{chi2017optimal}.
It is
traditionally assumed that $E\{M(Z)\}=1$, 
but that will be relaxed below. 

A reformulation of $\pi_I$ and the expected reinsurer 
surplus taken from \cite{erik2019}
will be needed later. 
Suppose $U=F(X)$ is the uniform under $X$. Then
\begin{displaymath}
\pi_I=E\{I(X)M(Z)\}=E\{E\{I(X)M(Z)|U\}\}=E\{I(X)E\{M(Z)|U\}\},
\end{displaymath}
with the last identity being due to $X=F^{-1}(U)$ having been fixed by $U$.
Hence
\begin{equation}
\pi_I=E\{I(X)W\{F(X)\}\}
\quad\mbox{where}\quad W(u)=E\{M(Z)|u\}
\label{reprem}
\end{equation}
which implies that the
reinsurer
expected surplus becomes
\begin{displaymath}
\pi(I)-E\{I(X)\}=\int_0^\infty(W\{F(x)\}-1)I(x)dF(x).
\end{displaymath}
Introduce
\begin{equation}
K(u)=\int_{u}^1\{W(v)-1\}dv,\hspace*{1cm}0\leq u\leq 1
\label{e213}
\end{equation}
and note that $dK(u)/du=-(W(u)-1)$ so that 
integration by parts yields
\begin{equation}
\pi(I)-E\{I(X)\}=\int_0^\infty K\{F(t)\} dI(t).
\label{e212}
\end{equation}
\subsection{Properties of the $\bf K$-function.}\label{subsec:2.3}

How $K(u)$ varies will provide useful information about the optimum
reinsurance functions in Section \ref{subsec:2.5}. It is reasonable to assume 
as in \cite{erik2019} that
$W(u)$ is an increasing function of $u$ which
is a form of positive dependence between $X$ and $M(Z)$. Then
$K(u)$ either increases to a maximum before
decreasing to $W(1)=0$ or decreases everywhere. Of
particular interest are the values at $u=0$ and $u=1-\epsilon$.
First note that~(\ref{reprem}) and~(\ref{e213}) yield
\begin{equation}
K(0)=E\{M(Z)\}-1
\label{ko}
\end{equation} 
so that $K(0)=0$ if $E\{M(Z)\}=1$. The latter
is a common assumption in actuarial literature which goes back to 
\cite{buhlmann1980economic},
yet it will be suggested in the next section that
$E\{M(Z)\}$ may well  be larger. It is in either case easy to verify that
\begin{equation}
K(u)\geq 0,
\label{ke}
\end{equation}
if $W(u)$ is increasing in  $u$. Suppose 
$W(u_m)=1$ which implies that
$W(u)\leq 1$ for $u\leq u_m$ and 
$W(u)\geq 1$ for $u>u_m$.
This means that $K(u)\geq 0$ for $u>u_m$ since the integrand in~(\ref{e213})
is positive 
everywhere while
\begin{displaymath}
K(u)=K(0)-\int_0^u(W(v)-1),
\end{displaymath}
and the integrand on the right is negative when $u\leq u_m$ so that again 
$K(u)\geq 0$.

\subsection{Criteria for optimization}\label{subsec:2.4}
Many contributors to reinsurance optimum theory work with an expected 
utility function. If ${\cal U}(y)$ is the utility of wealth $y$, the aim 
is to select $I(X)$
so that
\begin{equation}
{\cal C}_I=E({\cal U}\{R_I(X)\})\label{equ:utlity}
\end{equation}
is maximized; 
consult \cite{arrow1963,Kaluszka2008} and \cite{GUERRA2008529}.
Another popular approach is through the risk-adjusted surplus of the 
reinsurer. This is a
Lagrangian set-up of the form
\begin{equation}
{\cal C}_I=G_I-\lambda\rho_{R_I},
\label{e29}
\end{equation}
where $\lambda>0$ is a coefficient pricing risk; see \cite{balbas2009optimal,tan2011optimality,Jiang_2017} and \cite{Weng2017}.  
If~(\ref{e29}) 
is maximized for given values of the coefficient $\lambda >0$, 
the resulting solutions
define an efficient frontier of the Markowitz type with the minimum risk
$\rho_I$ obtainable  for a  given value of the expected surplus 
$G_I$. 

A related criterion
is to  minimize risk over expected
surplus so that
\begin{equation}
{\cal C}_I=\frac{\rho_{R_I}}{G_I}.
\label{e210}
\end{equation}
The resulting solution is also located
on the Markowitz frontier and corresponds to a certain $\lambda$ and a certain 
view on risk. To see this suppose $I$ and  $I_\lambda$
minimize~(\ref{e210}) and~(\ref{e29})
with $G_{I}$ and   $G_{I_\lambda}$ their expected gains.
If it is possible to select $\lambda$ so that $G_{I}=G_{I_\lambda}$, then
$I$ must minimize~(\ref{e210}) and $I_\lambda$~(\ref{e29}).

\subsection{One-layer contracts and optimality}\label{subsec:2.5}
One-layer contracts are defined mathematically as
\begin{equation}
I_{\bf a}(x)=\max(x-a_1,0)-\max(x-a_2,0)
\label{eia}
\end{equation}
with ${\bf a}=(a_1,a_2)^T$ 
a vector. 
The rest of the paper will examine the loss in
effectiveness when  $a_1$ and $a_2$ are calculated
from estimated parameters. This is relevant because
such contracts are often optimal
or at least not too far from that, and
the discussion will also throw light on
why VaR and CVaR based criteria
behave so differently with respect to   
estimation error. 

The optimum $I(X)$ under the risk-adjusted surplus~(\ref{e29}) 
was identified by  \cite{chi2017optimal} who established
a link to certain
functions $\psi_v(x)$ for the VaR risk measure and $\psi_c(x)$ for CVaR. 
Indeed, let  $x_\epsilon$ be the $1-\epsilon$ percentile 
for $X$ and ${\cal I}(B)$ the indicator function for the event $B$. Then
\begin{equation}
 \psi_v(x)=-K\{F(x)\}+(\lambda+\beta){\cal I}(x\leq x_\epsilon)
\label{psiv}
\end{equation}
and
\begin{equation}
 \psi_c(x)=\psi_v(x)+
(\lambda+\beta){\cal I}(x> x_\epsilon)\frac{1-F(x)}{\epsilon}.
\label{psic}
\end{equation}
With $\psi(x)$ either of $\psi_v(x)$ or $\psi_c(x)$ 
the optimum 
reinsurance function is
\begin{equation}
 I(x)=\int_0^x {\cal I}_{\psi(t)>0}\,dt,
\label{optre}
\end{equation}
which is a
multi-layer reinsurance contract with the number of layers  
depending on how many times $\psi(x)$ crosses $0$. 
It  was argued in \cite{erik2019}  
that the optimum solutions under the conditions in Section \ref{subsec:2.3}
are of the form 
\begin{equation}
I(x)=I_{\bf a}(x)+I_{\bf b}(x)
\label{eiab}
\end{equation}
with ${\bf b}=(0,b_2)^T$ and $b_2\leq a_1$. The prerequisite
for a $b$-layer starting at the origin is 
\begin{displaymath}
\psi(0)=-(E\{M(Z)\}-1)+\lambda+\beta >0,
\end{displaymath}
which is always satisfied when $E\{M(Z)\}=1$, but not when
$E\{M(Z)\}>1$. 
An argument for the 
stronger condition
$E\{M(Z)\}>1+\gamma$ has been put forward
in  \cite{erik2019}. Now
$ \psi(0)<0$ if $\lambda<\gamma-\beta$ with~(\ref{eiab}) reducing to a 
single layer if
the price on risk is smaller than
the loading minus the cost of capital.

These results have impact on the 
upper cut-off point $a_2$ too. Since $K\{F(x_\epsilon)\}=K(1-\epsilon)$,
it follows from~(\ref{psiv}) that
\begin{displaymath}
\psi_v(x_\epsilon)=-K(1-\epsilon)+\beta+\lambda
\end{displaymath}
while $\psi_v(x)=-K\{F(x)\}\leq 0$ if $x>x_\epsilon$.
Reinsurance layers thus do not extend beyond $x_\epsilon$ while there
is a change of sign there if $K(1-\epsilon)<\beta+\lambda$. This appears 
most common in practice, and gives $a_2=x_\epsilon$ as an optimal upper limit
when VaR is the risk measure. That changes with CVaR where the form of the
function $\psi_c(x)$ in~(\ref{psic}) shows that the optimum $a_2>x_\epsilon$.
This difference
has profound impact of the asymptotic theory in the next section
since criteria functions based on CVaR become smooth with 
second order derivatives
at the optimum points whereas there is a singularity at $a_2=x_\epsilon$
for VaR.

\section{Degradation }\label{sec3}

\subsection{ Formulation}\label{subsec:form}
In practice the distribution of $X$ depends on unknown parameters,
for example the claim intensity $\mu$ and expectation $\xi$
and shape parameter $\alpha$ of the claim severities, but we have only access to
estimated quantities which means that the reinsurance solutions are some 
distance from the real optimum. To put the problem in mathematical form
let the unknown parameters hiding under the distribution function 
$F(x;\B\theta)$
be a vector
$\B\theta=(\theta_1,\dots,\theta_{n_\theta})^T$
and consider  
some class
of reinsurance treaties defined by varying  
${\bf a}=(a_1,\dots,a_{n_a})^T$. 
It will be convenient to rewrite the
criterion ${\cal C}_I$ in~(\ref{e210}) as $C({\bf a},\B \theta)$
with the vector ${\bf a}$ defining the class
of reinsurance arrangements under consideration and  $\B \theta$
as the parameters under which it has been calculated.

The problem is that we do not have access to the true parameter vector
{\boldmath $\theta$}, only an estimated one
$\hat{\B \theta}$.
Suppose $\bf a$ and  $\hat{\bf a}$ define optimal reinsurance contracts
under $\B\theta$ and $\hat{\B\theta}$,
then
\begin{equation}
C({\bf a},\B\theta)=\min_{\bf b} 
C({\bf b}, \B\theta)
\quad \mbox{and}\quad
C(\hat{\bf a},\hat{\B\theta})=\min_{\bf b} 
C({\bf b},\hat{\B\theta}),
\label{e31}
\end{equation}
and we are interested in the difference
\begin{equation}
D(\B\theta)=C(\hat{\bf a},\B\theta)-C(\bf a,\B\theta),
\label{e32}
\end{equation}
where $C(\hat{\bf a},\B\theta)$ evaluates how well the 
optimal coefficient $\hat{\bf a}$ obtained under the estimate 
vector $\hat{\B\theta}$ works when 
$\bm{\theta}$ is the true one. Note that $D(\B\theta)\geq 0$, 
and the question is how much estimation error 
has made it grow.
\subsection{The bootstrap}\label{subsec:bootstrap}
One approach is through the bootstrap which yields
the mean and variance  and even the distribution  of 
$D(\B\theta)$. 
This means that the historical data is simulated from
the estimate 
which is then
re-estimated as (say) $\hat{\B\theta}^*$  and an alternative
optimal reinsurance treaty $\hat{\bf a}^*$ calculated. The 
distribution of $D(\B\theta)$ is then identified with 
\begin{equation}
D(\hat{\B\theta})=C(\hat{\bf a}^*,
\hat{\B\theta})-
C(\hat{\bf a},\hat{\B\theta}),
\label{e33}
\end{equation}
which can be examined by repeating the simulations $50$ or $100$ times.
In practice this amounts to nested bootstrapping since the criterion 
$C({\hat {\bf a}},\hat{\B\theta})$ is typically
computed by Monte Carlo. The  bootstrap approach is used
in Section \ref{sec:numercial1}.
\subsection{Asymptotics for smooth criteria}
\label{subsec:asy_smooth}
Many of the criteria used in theory of optimal reinsurance are smooth
functions of the coefficient vector $\B a$ in the sense that 
they are twice differentiable with respect to $\B a$. That will 
normally be the case when the risk measure  in~(\ref{e210}) 
is convex as in 
\cite{cheung2014optimal} and \cite{GAJEK2004227} or when 
$C({\bf a},\B\theta)$ in \eqref{equ:utlity} is minus the expectation of a 
utility
function  ${\cal U}(X)$ of the cedent.
It may also apply to CVaR-based criteria since the objective function is smooth
at the optimal coefficient vector $\bf a$, as was remarked in Section \ref{subsec:2.5},
and Proposition \ref{prop3.1} below may be valid for CVaR too.
Theoretical insight 
into how much parameter error
degrades optima
can  under these circumstances 
be gained through standard
asymptotics by letting the number of observations $n$
behind the estimate $\hat{\B\theta}$ become infinite.
Usually $\hat{\B\theta}$ then
becomes Gaussian with mean {$\B\theta$} 
and some covariance matrix $\Sigma/n$. The precise formulation is
\begin{equation}
\sqrt{n}(\hat{\B\theta}-\B\theta)
\stackrel{d}{\longrightarrow} {\bf N}
\quad\mbox{where}\quad {\bf N}\sim N(0,\Sigma)
\label{e34}
\end{equation}
with $\Sigma$ depending on {$\B\theta$}.
Let $C^{aa}=(c^{aa}_{ij})$ and  $C^{a\theta}=(c^{a\theta}_{ij})$ where
\begin{equation}
c_{ij}^{aa}=\frac{\partial^2C}{\partial a_i\partial a_j}
\quad\mbox{and}\quad
c_{ij}^{a\theta}=\frac{\partial^2C}{\partial a_i\partial \theta_j}
\hspace*{1cm}i,j=1,\dots,n_\theta
\label{e35}
\end{equation}
be second order
derivative matrices of  $C({\B a},\B \theta)$
There is then the following proposition.
\\
\begin{prop}\label{prop3.1}
	If $\hat{\B\theta}$
	is asymptotically Gaussian as in~(\ref{e34}) and
	$C({\bf a},\hat{\B\theta})$
	twice differentiable in $\bf a$ and
	{$ \B\theta$}, then as $n\rightarrow \infty$
	\begin{equation}
	nD(\mbox{\boldmath $\theta$})
	\stackrel{d}{\longrightarrow} {\bm N}^TQ{\bm N}
	\quad \mbox{where}\quad
	Q=\frac{1}{2}(C^{a\theta})^T(C^{aa})^{-1}(C^{a\theta}).
	\label{e36}
	\end{equation}
\end{prop}

The asymptotic distribution of
$D(\B\theta)$ is thus a Gaussian quadratic form,
consult Appendix \ref{proofA1} for the proof.
Mean and standard deviation in the asymptotic distribution 
are calculated on p. 424 in \cite{provost1992} and become
\begin{equation}
E\{D(\B\theta)\}=\frac{1}{n}\mbox{tr}(Q\Sigma)+o(1/n) 
\quad\mbox{and}\quad
\mbox{sd}\{D(\B\theta)\}=\frac{1}{n}
\sqrt{2\mbox{tr}(Q\Sigma Q\Sigma)}+o(1/n),
\label{e37}
\end{equation}
where $o(1/n)$ represents quantities for which
$o(1/n)\rightarrow 0$ as $n\rightarrow\infty$.
The operator tr is the trace of a matrix 
(the sum of its diagonal elements).

What is lost by not knowing $\theta$ is of order $1/n$. The 
bias $\E\{D(\B\theta)\}$ is always positive. 
This is not immediate 
from~(\ref{e37})
left, but the fact that $Q$ in~(\ref{e36}) right is positive definite shows that
it must be so. Practical  calculation requires the second order derivatives
of $C({\bf a},\B\theta)$ which must be carried out 
numerically.

\subsection{Asymptotics for non-smooth criteria}
\label{subsec:asy_nonsmooth}
The argument leading to Proposition \ref{prop3.1}
is based on an ordinary Taylor expansion and
doesn't work for the VaR criterion
which is  not differentiable at the optimum point $a_2=x_\epsilon(\B\theta)$. 
Consider the one-layer contract~(\ref{eia}) known from
Section \ref{subsec:2.5} to be optimal or close to that
under a wide class of 
reinsurance premium principles. To formulate the asymptotic result
we need the gradient vector of 
the
$1-\epsilon$ percentile $x_\epsilon(\B \theta)$ with respect to $\B \theta$; i.e.
\begin{equation}
{\bf g}=(g_1,\dots,g_{n_\theta})^T,
\quad
g_i=\frac{\partial x_\epsilon(\mbox{\boldmath $\theta$})}
{\partial \theta_i},\quad i=1,\ldots,n_\theta,
\label{e38}
\end{equation}
and also
\begin{equation}
V=\frac{{\bf g}^T{\bf N}}{\sqrt{{\bf g}^T\Sigma{\bf g}}}
\label{e38b}
\end{equation}
with $\bf N$  the same Gaussian vector as in the preceding section
which implies that $V$ is Gaussian $(0,1)$.
We also need the coefficients
\begin{equation}
h_1({\bf a},\B\theta)=\{1+\beta C({\bf a},\B\theta)\} C({\bf a},\B\theta)/a_1
\quad\mbox{and}\quad
h_2({\bf a},\B\theta)=C({\bf a},\B\theta)^2K(1-\epsilon)/a_1.
\label{e38a}
\end{equation}
\begin{prop}\label{prop3.2}
	If $\hat{\B\theta}$
	is asymptotically Gaussian as in~(\ref{e34}),
	and $C({\bf a},\B\theta)$ is the VaR over
	expected gain criterion, then as $n\rightarrow \infty$
	\begin{equation}
	\sqrt{n}D(\B\theta)
	\stackrel{d}{\longrightarrow} 
        \{h_1({\bf a},\B\theta)(-V)_++
        h_2({\bf a},\B\theta)V\}\sqrt{{\bf g}^T\Sigma{\bf g}},
	\label{e39}
	\end{equation}
where $V$ is Normal (0,1).
\end{prop}

Note that degradation now shrinks at the rate rate $1/\sqrt{n}$
instead of $1/n$ as in Proposition \ref{prop3.1}. 
Expectation and variance in the asympotic distribution
become
\begin{equation}
E\{D(\B\theta)\}=\frac{1}{\sqrt{2\pi n}}
h_1({\bf a},\B\theta)\sqrt{{\bf g}^T\Sigma{\bf g}}
\label{e310}
\end{equation}
and
\begin{equation}
\Var\{D(\B\theta)\}=\frac{1}{4n}\{(1-1/\pi)
h_1({\bf a},\B\theta)^2+2h_1({\bf a},\B\theta)
h_2({\bf a},\B\theta)+2h_2({\bf a},\B\theta)^2\}
{\bf g}^T\Sigma{\bf g}.
\label{e310a}
\end{equation}
All these results are verified in Appendix \ref{proofA2}.

\section{Numerical study}\label{sec:numercial1}
\subsection{Candidate models}\label{subsec:model}
In non-life insurance, the stochastic model for the total loss $X$ is typically split into separate models for the claim numbers $N$ and the individual losses $Y_i$, called the claim frequency and the claim severity distribution, respectively. The collective risk model for $X$ is then given as
\begin{displaymath}
X=\sum\limits_{i=1}^{N}Y_i.
\end{displaymath}
The claim severities are commonly assumed identically distributed and independent of each other and of the claim number \citep{kaas2001, klugman2012loss}.
In this study, we restrict our attention to the choice of the claim severity distributions, fixing the claim frequency distribution at the Poisson distribution  with fixed intensity $\mu$ for all policies. Three classic right-skewed distributions are considered for claim severities, namely, the Gamma, the Lognormal and the Pareto. The Gamma distribution used throughout this paper is parameterized as
\[
f(y)= \frac{y^{\alpha-1}e^{-y/\beta}}{(\beta)^{\alpha}\Gamma(\alpha)}, \quad y > 0
\]
with a shape parameter $\alpha$ and a scale parameter $\beta$ so that $\E(y)=\alpha\beta$.  The Pareto distribution, which is also called the Pareto type II or the Lomax distribution, has two parameters, the shape $\alpha$ and the scale $\beta$. The pdf is given by
\begin{displaymath}
f(y) = \frac{\alpha/\beta}{(1+y/\beta)^{\alpha+1}},\quad y > 0.
\end{displaymath}
Additionally, the Gaussian distribution $N(\E(X),\sd(X))$ is used for approximating the total loss distribution, where 
\begin{equation*}
\E(X)=\E(N) \E(Y_i)\quad \mbox{and} \quad  \sd(X)=\sqrt{\Var(N)\E(Y_i)^2+\E(N)\Var(Y_i)}.
\end{equation*}

%In this subsection, a simulation study is performed to examine the degradation of optimal reinsurance when the sample size varies. In particular, we investigate how many data are needed to fit the parameters and distributions. The parameter settings and simulation procedure for this study are provided in Section \ref{sec:simluation_set}, and results of the bootstrap in Section \ref{subsec:bootstrap} are shown in Section \ref{sec:uni_error}.

\subsection{Parameter settings}
\label{sec:simluation_set}
Table \ref{tab:parameters} summarizes the choice of parameter values for the claim severity distributions with corresponding skewness coefficients and $99\%$ and $99.5\%$ reserves ($\epsilon=0.01$ and $0.005$, respectively). These parameter values are calibrated so they have a common mean of (approximately) $10$ and a standard deviation of $15$. The claim frequency distribution follows Poisson$(J\mu T)$, where $J=1000$ is the number of policies and $\mu= 0.05$ the claim intensity during one year ($T=1$). The loading factors are  $\gamma =0.1$ and $\gamma_r=0.2$, respectively. 
%The Gaussian approximation method is also used to represent the total loss distribution with corresponding mean and standard deviation.
%$\mu$ is assumed stochastic, namely, $\mu \sim \mathrm{Gamma}(25,0.05)$ with a mean of $0.05$ and a standard deviation of $0.01$. 
%As mentioned in Section \ref{subsec:claimsize}, we assume the total claim loss follows a Gaussian distribution with a mean of $500$ and a standard deviation of (approximately) $127$.

The total claim amount is simulated by means of Monte Carlo for all sorts of models and stored in the computer prior to search for the best coefficients. The number of simulation is $m=1000000$ and the procedure is described in Algorithm \ref{alg:total_loss} in Appendix \ref{appen:b}. The sample size is varied between $n=5000, 500, 50$, representing a large, medium and small sample size, respectively. Then~\eqref{e210} is minimized by R function $\verb|optim|$ using the Nelder–Mead method \citep{Nelder}. This method seems more natural to use for the VaR-based criteria than the Quasi-Newton method \citep{wright1999} since it does not require the derivatives of the objective function. Also, A bisection based optimization algorithm is performed in Fortran to verify the results. 

%A robust numeric optimisation solution requires appropriate algorithm and sound initial values. In the case of single line of business, $a+b=q_\epsilon$ if the reinsurance premium is computed based on expected premium principle. Thus there is only one free parameter left in the problem and could be solved by $\verb|optimize|$ function numerically in \verb|R| language.

%As there are no analytic results when there are multiple lines of business, a numeric algorithm with appropriate initial values is vital to search for the optima.
%A sound choice of initial values to start with might be the sub-optimal values of $a_i$ and $b_i$ as if there were single line of business. It will also facilitate us to investigate how the sub-optima changes to reach the joint optimality. Furthermore, the Nelder–Mead method may be more natural than Quasi-Newton method in our simulation study, but there is no guarantee on finding the true local optima. A bisection based optimisation algorithm is performed in Fortran to verify the results. 

\begin{table}[!t]
	\small
	\centering
	\begin{tabular}{ccccc}
		\hline\hline
		Model & Parameters  & Skewness coef.&$99\%$ Res.&$99.5\%$ Res.\\
		\hline
		Gamma$(\alpha,\beta)$ &$(0.44,22.50)$ &$3.00$&938.79&995.07\\
		Lognormal$(\sigma,\xi)$ &$(1.71,1.09) $&$7.88$&866.56&925.99\\
		Pareto$(\alpha,\beta)$ & $(3.60, 26.00)$  &$5.78$&955.23&1024.67\\
		\hline\hline
	\end{tabular}
	\caption{Parameter values and properties for claim severity distributions}
	\label{tab:parameters}
\end{table}

\subsection{Results}\label{sec:uni_error}
The examples in this section  
have been run with
\begin{equation}
M(Z)=(1+\gamma_r)e^{\omega Z}/\E(e^{\omega Z}),
\end{equation}
where $\omega \geq 0$ is called the tilting parameter. By letting $Z=I(X)$, $\pi_I$ in \eqref{e25} becomes
\begin{equation}
\pi_I=(1+\gamma_r)E\{I(X)e^{\omega I(X)}\}/\E\{e^{\omega I(X)}\}.
\label{equ:esscher}
\end{equation}
This formulation is similar to the Esscher 
premium principle but is modified such that 
the expected premium principle is included as a 
special case, namely, when $\omega = 0$. In the rest of the 
paper \eqref{equ:esscher} is referred as the mixed Esscher premium principle.

\begin{figure}[!htp]
	\centering	\includegraphics[width=10cm,height=10cm,keepaspectratio]{{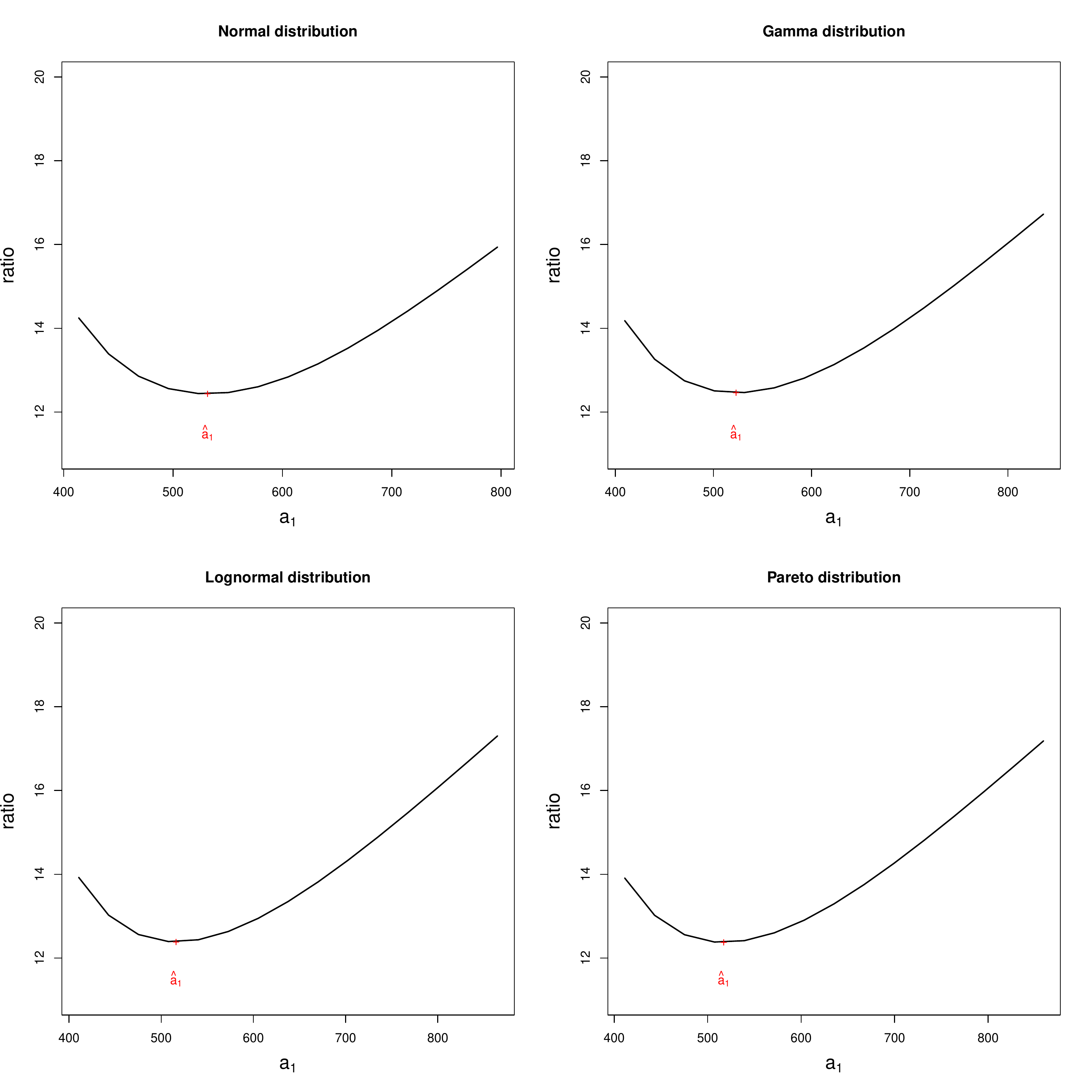}}
	\caption{The ratio $C({\bf a},\B\theta)$ as a function of the lower bound $a_1$ under the expected premium principle.}
	\label{fig:basic_a}
\end{figure}

Table \ref{tab:single_line1} displays the optimal coefficients and corresponding minimized ratio for the VaR-based criterion under different premium principles in an error-free environment. The results are computed under the three candidate claim severity distributions, with the Gaussian approximation as a comparison. For both of the premium principles, the optimized ratio of risk over surplus seems stable regardless of the choice of the claim severity distribution, while the upper limit $a_2$ varies a lot since it depends on the heaviness of the claim severity distribution. Also, the lower bound $a_1$ and ratio are much higher for the mixed Esscher principle, which demonstrate that the cedent has the incentive to transfer more loss if the reinsurance premium is less costly. 
The curves in Figure \ref{fig:basic_a} illustrate the change of the ratio as a function of the lower bound $a_1$ with varying claim severity distributions and the Gaussian approximation. It again verifies the choice of the claim severity distributions does not make too much difference on deciding the optimal ratio. Table \ref{tab:basic_re} and \ref{tab:basic_omega} show the optimal values under different reinsurance risk loading $\gamma_r$ and tilting parameter $\omega$, respectively. %The results also verify that the more expensive the reinsurance is, the less loss will be ceded to the reinsurer. 
In all examples, the 
cost of capital rate $\beta$ is not taken into account, but they can be incorporated easily. 
%The results for CVaR are quite similar to VaR and shown in Table~\ref{tab:single_line1_cvar}. %(POSSIBLE REASON?)
\begin{table}[!htp]
	\centering
	\small
	\begin{tabular}{cccccccc}
		\hline\hline
		\multirow{2}{*}{Model}& \multicolumn{3}{c}{Expected Prem.}&& \multicolumn{3}{c}{Mixed Esscher Prem.}\\
		\cline{2-4}
		\cline{6-8}
		&$a_1$ & $ a_2- a_1$ & $C({\bf a},\B\theta)$&&$ a_1$ & $a_2- a_1$ & $C({\bf a},\B\theta)$\\
		\hline
		Gaussian&531.5&264.4&12.43&&598.6&197.3&  13.33\\
		Gamma&523.3& 312.7& 12.46&&605.0&231.0& 13.64\\
		Lognormal&516.7&349.9&12.39&&604.6& 262.0&  13.79\\
		Pareto&516.9& 344.1&  12.37&&602.1& 259.0&  13.71\\
		\hline\hline
	\end{tabular}
	\caption{The optimal results for the VaR-based criterion for both the expected and the mixed Esscher premium principles when $\epsilon=0.01$, $\gamma=0.1$, $\gamma_r=0.2$, $\omega=0.001$ and $ \beta=0$.}
	\label{tab:single_line1}
\end{table}

\begin{sidewaystable}[!h]
	\small
	\centering
	\begin{tabular}{ccccccccccccc}
		\hline\hline
		&\multicolumn{3}{c}{\textbf{Gaussian}}&\multicolumn{3}{c}{\textbf{Gamma}}&\multicolumn{3}{c}{\textbf{Lognormal}}&\multicolumn{3}{c}{\textbf{Pareto}}\\
		$\gamma_r$&$\hat a_1$ & $\hat a_2-\hat a_1$ & $C(\hat{\bf a},\B\theta)$&$\hat a_1$ & $\hat a_2-\hat a_1$ & $C(\hat{\bf a},\B\theta)$&$\hat a_1$ & $\hat a_2-\hat a_1$ & $C(\hat{\bf a},\B\theta)$&$\hat a_1$ & $\hat a_2-\hat a_1$ & $C(\hat{\bf a},\B\theta)$\\
		\hline
		$0.2$ &531.5&264.4 & 12.43&523.3& 312.7&  12.46&516.7&349.9&12.39&516.9& 344.1&  12.37\\
		$0.3$ &584.7& 211.2& 13.19&580.9& 255.1&  13.34&573.7& 292.9&13.33& 573.4& 287.7& 13.28\\
		$0.4$ & 615.0& 180.9& 13.65&614.9& 221.1& 13.88&608.7& 257.9&13.92& 607.5& 253.5& 13.87\\
		$0.5$ &635.7& 160.2&13.97&638.6& 197.4& 14.26& 633.9& 232.7& 14.36& 632.1&228.9&  14.29\\
		$0.6$ &651.5& 144.4&  14.21&656.9&179.1& 14.56&653.8& 212.7&  14.70&651.4& 209.7& 14.62 \\
		$0.7$ &664.0& 131.9&14.40&671.6& 164.4& 14.80 &670.0& 196.6&14.97& 667.1& 193.9& 14.89\\   
		\hline\hline
	\end{tabular}
	\caption{The optimal reinsurance coefficients for different risk loadings $\gamma_r$.}
	\label{tab:basic_re}
	\vspace{2\baselineskip}
	\centering
	\begin{tabular}{ccccccccccccc}
		\hline\hline
		&\multicolumn{3}{c}{\textbf{Gaussian}}&\multicolumn{3}{c}{\textbf{Gamma}}&\multicolumn{3}{c}{\textbf{Lognormal}}&\multicolumn{3}{c}{\textbf{Pareto}}\\
		$\omega$&$\hat a_1$ & $\hat a_2-\hat a_1$ & $C(\hat{\bf a},\B\theta)$&$\hat a_1$ & $\hat a_2-\hat a_1$ & $C(\hat{\bf a},\B\theta)$&$\hat a_1$ & $\hat a_2-\hat a_1$ & $C(\hat{\bf a},\B\theta)$&$\hat a_1$ & $\hat a_2-\hat a_1$ & $C(\hat{\bf a},\B\theta)$\\
		\hline
		\hline
		$0.001$ &598.5& 197.4&  13.33&605.0& 231.0& 13.64&604.6& 262.0&  13.79&602.1& 259.0&13.71\\
		$0.002$ &633.3& 162.6&  13.80&648.3& 187.7&  14.26&654.7& 211.8&14.54&650.8& 210.2& 14.45\\
		$0.003$ &656.0& 139.9&  14.11&676.5& 159.5&  14.66&688.1& 178.4&15.03&683.5& 177.5&  14.93\\
		$0.004$ &672.4& 123.5& 14.34&697.0& 139.0& 14.94&712.6& 154.0& 15.37&707.4& 153.6& 15.27\\
		$0.005$ &685.0&110.9&14.51&712.7& 123.3& 15.16&731.2& 135.4&15.63&726.0& 135.1& 15.53\\
		$0.006$ &695.2&100.7& 14.65&725.1& 110.9&  15.32&745.9& 120.6&15.83&740.5& 120.6&15.72\\   
		\hline\hline
	\end{tabular}
	\caption{The optimal reinsurance coefficients for different titling parameter $\omega$.}
	\label{tab:basic_omega}
\end{sidewaystable}

Then the degradation of the optimal ratio due to parameter error is examined through the bootstrap method introduced in Section \ref{subsec:bootstrap}. Table \ref{tab:boot_expt} and \ref{tab:boot_esscher} illustrate how far the optimum now have moved when the number of historical observations $n$ varies with $\pi_I$ computed by the expected premium and the mixed Esscher premium principle, respectively. It can be seen that the degradation $D(\hat{\B \theta})$ shrinks approximately at the rate of $O(1/\sqrt{n})$ for all claim severity distributions, except the case when $n=50$. It makes sense since the asymptotics works for large samples. More importantly, it is not rational for the cedent to buy reinsurance when only few data are available, as the error is rather large in this case. %Further, given the mean and standard deviation of the claim severities, the estimated ratio $\hat C_I$ is virtually the same whatever the claim severity distribution is, suggesting that the Gaussian approximation as an alternative is fairly acceptable for medium and large samples. 
It is also possible to gain some insights on the number of samples needed through the degradation analysis. As an example, Table \ref{tab:no_of_claim} reports the amount of data that are required under different significance levels based on the root mean squared error (RMSE).
The Gamma distribution is examined with different standard deviations, representing a light tail and comparatively heavy tail. The required sample size at each level further verifies that the degree of degradation is of order $1/\sqrt n$ and shows that it depends on the heaviness of the underlying model.

\begin{table}[H]
	\centering
	\small
	\begin{tabular}{llcccccc}
		\hline
		\hline
		Model & $n$& $\E(\hat{a}_1^*)$ &$\E(\hat{a}_2^*)$ & $\E(C(\hat{{\bf a}}^*, \hat{\B{\theta}}))$&$C(\hat{{\bf a}}, \hat{\B{\theta}})$ &$\E[D(\hat{\B{\theta}})]$&$\sd[D(\hat{\bm{\theta}})]$\\
		\hline
		\multirow{3}{*}{Gamma}& 5000&522.9& 835.7&12.71 &12.46&0.255 &0.364\\
		&500& 520.4& 832.4& 13.35 &12.40&0.893&1.212  \\
		&50& 493.6& 790.5&17.66 &12.49&5.207& 5.783\\
		\hline
		\multirow{3}{*}{Lognormal} & 5000&513.7&861.5&12.68 &12.39&0.289& 0.418\\
		&500&518.1& 872.7& 13.17 &12.23&0.786&1.012\\
		&50&538.8& 919.2&15.47&12.14&3.081&3.982\\
		\hline
		\multirow{3}{*}{Pareto} & 5000&516.1& 853.4&  12.75 &12.33&0.378&0.463\\
		&500&510.6& 849.4&13.56&12.66&1.187& 1.423\\
		&50&525.8& 998.1& 16.35&13.13&3.980&4.892\\
		\hline\hline
	\end{tabular}
	\caption{Nested bootstrapping errors of  $D(\hat{\B{\theta}})$  under VaR with varying claim size distribution when $\pi_I$ is calculated by the expected premium principle.}
	\label{tab:boot_expt}
\end{table}

\begin{table}[H]
	\centering
	\small
	\begin{tabular}{llcccccc}
		\hline
		\hline
		Model & $n$& $\E(\hat{a}_1^*)$ &$\E(\hat{a}_2^*)$ & $\E(C(\hat{{\bf a}}^*, \hat{\B{\theta}}))$&$C(\hat{{\bf a }}, \hat{\B{\theta}})$ &$\E[D(\hat{\B{\theta}})]$&$\sd[D(\hat{\bm{\theta}})]$\\
		\hline
		\multirow{3}{*}{Gamma} & 5000& 614.2&848.2&13.85&13.64&0.208&0.259\\
		&500&654.9&896.0&14.39&13.59&0.803&1.008 \\
		&50& 694.4&945.7&17.49 &13.79&3.695&4.055\\
		\hline
		\multirow{3}{*}{Lognormal} & 5000&598.2&856.0&13.95&13.73&0.221&0.272\\
		&500&524.7&746.4&14.14&13.33&0.812&0.977\\
		&50&572.7&800.5&16.18&13.17&3.011&3.289\\
		\hline
		\multirow{3}{*}{Pareto} & 5000&572.5&814.4&13.86&13.56&0.298&0.334\\
		&500&683.3&1005.5&16.25&14.74&1.534&1.693\\
		&50&579.2&769.4&15.17&12.59&2.580&2.841 \\
		\hline\hline
	\end{tabular}
	\caption{Nested bootstrapping errors of  $D(\hat{\B{\theta}})$ under VaR with varying claim size distribution when $\pi_I$ is calculated by the mixed Esscher premium principle.}
	\label{tab:boot_esscher}
\end{table}

\begin{table}[H]
	\centering
	\small
	\begin{tabular}{lccccccc}
		\hline\hline
		\multirow{2}{*}{Model}&\multirow{2}{*}{Std}&\multicolumn{2}{c}{5\% RMSE}&\multicolumn{2}{c}{15\% RMSE}&\multicolumn{2}{c}{25\% RMSE}\\
		\cline{3-8}
		&&$n$&$\mathrm{RMSE}[D(\hat{\bm{\theta}})]$&$n$&$\mathrm{RMSE}[D(\hat{\bm{\theta}})]$&$n$&$\mathrm{RMSE}[D(\hat{\bm{\theta}})]$\\
		\hline
		\parbox{2cm}{Gamma\\(4, 2.5)}&5& 120000 &0.0494&11000&0.1536&5800&0.2469\\
		\parbox{2cm}{Gamma\\(0.44, 22.5)}&15&150000 & 0.0539&21000&0.1489&8800&0.2491\\
		%\hline
		%Lognormal&\\
		%Lognormal&\\
		%\hline
		%Pareto&\\
		%Pareto&\\
		\hline\hline
	\end{tabular}
	\caption{The number of observations needed for Gamma  distribution with different standard deviations when $\pi_I$ is calculated by the expected premium principle.}
	\label{tab:no_of_claim}
\end{table}

\section{A Bayesian approach}\label{sec:bayes}
\subsection{Method}
The Bayesian paradigm expresses prior belief about the parameters $\B \theta$ as a probability distribution, the so-called prior, which is updated on observing historical data $(n,\B y)$, where $n$ is the number of incidents and $\B y =(y_1,\ldots,y_n)$ their size. Now let $\B \theta= (\mu,\B\zeta)$, representing the parameters from the claim frequency and the claim severity distribution, respectively. The posterior distribution is via Bayes' rule 
\begin{equation}\label{bayesian}
p(\mu,\B\zeta|n,\B{y})\propto f(n,\B{y}|\mu,\B\zeta)p(\mu,\B\zeta),
\end{equation}
where $p(\mu, \B \zeta) $ is the prior density of $(\mu,\B \zeta)$ and $f(n,\B{y}|\mu, \B\zeta)$ the likelihood of the observations.  The symbol $\propto$ signifies that a normalising constant that does not depend on $(\mu,\B \zeta)$ has been omitted. In many applications $n$ is a realization of a Poisson variable with known exposure $A$, so that its parameter is $\mu A$. If claim frequency is stochastically independent of claim severity and the same applies to $(n,\B y)$, the posterior distribution of $(\mu,\B\zeta)$ boils down to
\begin{equation}
p(\mu,\B\zeta|n,\B{y})= p(\mu|n) p(\B\zeta|\B{y}),
\end{equation}
with $p(\mu|n)$ and  $p(\B\zeta|\B{y})$ the posterior distributions for $\mu$ and $\B\zeta$. This opens for another way of taking parameter error into account. As a basis for setting up the the reinsurance contract replace the former $f(x; \mu,\B\zeta)$ or rather its estimate $f(x; \hat{\mu},\hat{\B\zeta})$ with the so-called posterior predictive distribution
\begin{displaymath}\label{equ:pred}
p(x|n,\B{y})= \int p(x|\mu, \B \zeta) p(\mu|n) p(\B\zeta|\B{y})d\mu \;d\B\zeta,
\end{displaymath}
%where $x$ is an unknown observable. The predictive distribution can be viewed as a weighted average of the model $p(\B y|\mu,\B\theta)$. The weights are determined by the posterior distribution of $p(\mu,\B\theta|\B{y})$, meaning that the 
and parameter uncertainty is  incorporated automatically. 
%  Now it is a matter of how the parameter uncertainty is incorporated into the optimization problem defined in Section \ref{sec:basics}.  Point estimates $(\hat\mu,\hat{\B\theta})$ are adopted as an approximation for the parameters in the frequentist method 
The risk measure $\rho_I$ and the expected gain of the cedent $G_I$ earlier calculated under $(\hat\mu,\hat{\B\zeta})$, now depend on $p(x|n,\B y)$ in which the uncertainty of $(\mu,\B\zeta)$ is embedded. 

We need posterior quantities to express the risk over surplus ratio ${\cal C}_I$, for example
\begin{equation}
x_{\epsilon|n,\B y}=\mbox{VaR}_\epsilon(X|n,\B y) \quad
\mbox{and}\quad \pi_I(X|n, \B y)= E\{I(X)W\{F(X)\}|n,\B y\},
\end{equation}
where $I(\cdot)$ and $R_I(\cdot)$ are defined earlier, see Section \ref{sec:basics}. With VaR as a risk measure,  ${\cal C}_I$ in this posterior setting becomes
\begin{equation}\label{equ:5.4}
{\cal C}_I=\frac{x_{\epsilon|n,\B y}-I(x_{\epsilon|n,\B y})}{\gamma E(X|n,\B y)-\beta x_{\epsilon|n,\B y}-\pi_I(X|n, \B y)+E\{I(X|n, \B y)\}+\beta I(x_{\epsilon|n,\B y})},
\end{equation}
but the posterior density function $p(x|n,\B y)$ is complicated, and there is no closed form. Monte Carlo is a way around. For a single simulation $X^*$,  draw
\begin{equation}
\label{equ:5.6}
\mu^\star \sim p(\mu|n),\quad \B\zeta^\star \sim p(\B\zeta|\B{y})\quad \mbox{and} \quad X^\star \sim p(x|\mu^\star,\B\zeta^\star),
\end{equation}
and repeating $m$ times yields a posterior sample  $X^\star_1,\dots,X^\star_m$ depending on $(n,\B y)$. The symbol $^\star$ here marks for posteriors. The quantities in \eqref{equ:5.4} are replaced by their Monte Carlo analogues. Those are 
\[
x_{\epsilon|n,\B y}\approx X^\star_{(\epsilon m)}
\]
for the ordered sample $X^\star_{(1)}\geq \dots \geq X^\star_{(m)}$ \mbox{of} $X_1^\star,\dots,X_m^\star$, and 
\begin{align*}
&E(X|n,\B y) \approx \frac{1}{m}\sum\limits_{i=1}^{m}X^\star_i,\\
&I(X|n,\B y)\approx \sum\limits_{i=1}^{m}\max(X_i^\star-a_1,0)-\max(X_i^\star-a_2,0),\\
&\pi_I(X|n, \B y)\approx \frac{1}{m}\sum\limits_{i=1}^{m}I(X^\star_i)W\{F(X_i^\star)\}.
\end{align*}
The optimal coefficient $\hat {\bf a}_B= (\hat{a}_1,\hat{a}_2)$ under $p(x|n,\B y)$ can then be computed numerically with the corresponding value of $C_I$. Evaluation of the procedure is a problem of its own. The degradation in \eqref{e32} can still be written 
\begin{equation}
D(\mu, \B \zeta)=C(\hat{\bf a}_B; \mu, \B \zeta)-C({\bf a}; \mu, \B \zeta),
\label{bayes_D}
\end{equation}
but in a Bayesian model the true parameters $(\mu,\B\zeta)$ are random. One way to go about would then be to draw $(\mu,\B\zeta)$ from their priors, generate Monte Carlo historical data $(n,\B y)$ given $(\mu,\B\zeta)$ and then $X_1^\star,\dots,X^\star_m$. The entire procedure is summarized in Table \ref{tab:alg_bayes}.
%and the corresponding $D(\mu_1^\star,\B{\zeta}^\star_1),\dots, D(\mu_m^\star,\B{\zeta}^\star_m)$ of $D(\mu,\B{\zeta})$. Their mean and standard deviation approximate $\E\{D(\mu, \B \zeta)\}$ and $\sd\{D(\mu, \B \zeta)\}$ in the usual way. Note this is nested Monte Carlo with $m_b$ repetitions in the outer loop.

\begin{table}[H]
	\centering
	\small
	\begin{tabular}{l|l}
		\hline\hline
		Step&Procedure\\
		\hline
		1.& $\mu\sim p(\mu),\quad \B\zeta \sim p(\B\zeta)$\\
		2.&$n\sim \mbox{Piosson}(\mu A), \quad \B y\sim f(\B\zeta)$\\
		3.& $X_1,\dots, X_m, \quad\mbox{where}\quad X_i\sim p(x|\mu,\B{\zeta}),$ \quad see Algorithm \ref{alg:total_loss} for details\\
		4.& $\mu^\star_1,\dots, \mu^\star_m \sim p(\mu|n), \quad \B{\zeta}^\star_1,\dots,\B{\zeta}^\star_m \sim p(\B\zeta|\B y)$\\
		5.& $X^\star_1,\dots, X^\star_m, \quad\mbox{where}\quad X^\star_i\sim p(x|\mu^\star_i,\B{\zeta}^\star_i)$,\quad see Algorithm \ref{alg:bayes_crit} for details\\
		6.& $\hat{\bf a}_B=\underset{\bf{b}}{\argmin}\; C({\bf{b}};X^\star_1,\dots, X^\star_m)$\\
		7.& ${\bf a}=\underset{\bf{b}}{\argmin}\; C({\bf{b}};X_1,\dots, X_m)$\\
		8.& $D(\mu,\B{\zeta})= C({\hat{\bf a}_B};X_1,\dots, X_m)-C({\bf{a}};X_1,\dots, X_m)$\\
		\hline\hline
	\end{tabular}
\caption{Simulation procedure of the Bayesian method.}\label{tab:alg_bayes}
\end{table}
\noindent
Repeat the procedure $m_b$ times yields $m_b$ replications  of $D(\mu,\B{\zeta})$, and 
\begin{displaymath}
\E\{D(\mu, \B \zeta)\} \approx \frac{1}{m_b}\sum\limits_{i=1}^{m_b}D_i(\mu,\B{\zeta}),
\end{displaymath}
\begin{displaymath}
 \sd\{D(\mu, \B \zeta)\}\approx \sqrt{\frac{1}{m_b-1}\sum\limits_{i=1}^{m_b}(D_i(\mu,\B{\zeta})-\E\{D(\mu, \B \zeta)\})^2}.
\end{displaymath}
%Notwithstanding the bias $\E\{D(\mu, \B \theta)\}$ and standard deviation $\sd\{D(\mu, \B \theta)\}$ can be computed by replicating the procedure (say) $m_b$ times.

We have in the numerical study below followed a slightly different track in order to compare with the frequentist method. Instead of drawing $(\mu,\B\zeta)$, $(\mu,\B\zeta)$ is fixed as $(\mu_0,\B\zeta_0)$ with the prior placed around it. The degradation in~\eqref{bayes_D} now becomes
\begin{equation}
D(\mu_0, \B \zeta_0)=C(\hat{\bf a}_B; \mu_0, \B \zeta_0)-C({\bf a}_0; \mu_0, \B \zeta_0),
\end{equation}
where ${\bf a}_0$ are the optimum under $(\mu_0, \B \zeta_0)$. The difference is that $(\mu,\B\zeta)$ in Step 1 in Table \ref{tab:alg_bayes} is replaced by $(\mu_0, \B \zeta_0)$ and Monte Carlo historical data $(n,\B y)$ are generated given $(\mu_0,\B\zeta_0)$ each time. We then get a different version  of $\E\{D(\mu_0, \B \zeta_0)\}$ and  $\sd\{D(\mu_0, \B \zeta_0)\}$ by replicating the above procedure $m_b$ times. 

%The true parameters $(\mu_0,\B{\theta}_0)$ can be replaced by randomly sampling from their prior distributions each time and the distribution of $\hat{\bf a}_B$ is obtained by repeating $m_b$ times. This is the natural way of solving the optimization problem in the Bayesian context, but the difference can not be compared directly with the frequentist method. The two different methods are both explored here.

%If that is carried out $m_b$ times, yielding $m_b$ Monte Carlo versions $D(\mu_1^\star,\B{\theta}^\star_1),\dots, D(\mu_m^\star,\B{\theta}^\star_m) $, their mean and standard deviation approximate $\E\{D(\mu, \B \theta)\}$ and $\sd\{D(\mu, \B \theta)\}$.

\subsection{Implementation issues}\label{subsec:5.2}
The degree of prior knowledge can be expressed through informative or non-informative priors.  Among the informative ones conjugates are popular since the functional form of the posterior can be calculated with easy implementation in the computer. Non-informative priors are used to reflect minimal knowledge and there is no consensus as to how it should be constructed. Often used in scientific literature and included here is the Jeffreys prior,  which has the form 
\[p(\B\theta)\propto \sqrt{\det I(\B\theta)}\quad \mbox{with} \quad I(\B\theta)=-\E\{\frac{\partial^2}{\partial\B{\theta}^2}\log f(y,\B\theta)|\B\theta\}.\]
The Jeffreys priors for the claim frequency and candidate claim severity distributions are 
\begin{align*}
&\mbox{Poisson:}\quad p(\mu)\propto \sqrt{(1/\mu)},\\
&\mbox{Gamma:}\quad p(\alpha,\beta) \propto \sqrt{(\alpha\phi(\alpha)-1)}/\beta,\;\mbox{with}\;\phi(\alpha)=\frac{d^2}{d\alpha^2}\log\Gamma(\alpha),\\
&\mbox{Lognormal:}\quad p(\xi,\sigma^2)\propto\sigma^{-3},\\
&\mbox{Pareto:}\quad p(\alpha,\beta) \propto\Big(\beta (\alpha+1)\sqrt{(\alpha(\alpha+2))}\Big)^{-1},
\end{align*}
and corresponding posteriors can be written up to some unknown constant. Note for the Lognormal parameters a slightly different prior is used, which is  
\[
p(\xi,\sigma^2)\propto\sigma^{-2}.
\]
This prior corresponds with a flat prior on $\log(\sigma)$ and leads to a relatively simple posterior with closed form; that's why it is  quite often used in practice. 

Table \ref{tab:conju} lists  conjugate prior distributions for the Poisson claim frequency and various claim severity distributions. The conjugate prior typically allows the use of Gibbs sampling, which is less computationally intensive. For those parameters that do not have conjugates, we use the Gamma distribution as informative priors. In this case, the corresponding posterior distribution ends up as a non-standard and analytically intractable function. Metropolis-Hastings (MH) is a suitable method to generate the parameters. 
%It is also the conjugate prior for the reciprocal of variance parameter $(1/\sigma^2)$ in the Lognormal distribution and the reciprocal of scale parameter $(1/\beta)$ in the Gamma and Pareto distribution.
%inplementation part..
\begin{comment}

\begin{table}[!t]
	\centering
	\small
	\begin{tabular}{c|c}
		\hline\hline
		Model&Non-informative Prior\\
		\hline
		Pois.&$p(\mu)\propto \sqrt{(1/\mu)}$\\
		\hline
		Ga.&$p(\alpha,\beta) \propto \sqrt{(\alpha\phi(\alpha)-1)}/\beta,\;\mbox{with}\;\phi(\alpha)=\frac{d^2}{d\alpha^2}\log\Gamma(\alpha)$\\
		\hline
		L.N.&$p(\xi,\sigma^2)\propto\sigma^{-2}$\\
		\hline
		Pa.&$p(\alpha,\beta) \propto\Big(\beta (\alpha+1)\sqrt{(\alpha(\alpha+2))}\Big)^{-1}$\\
		\hline\hline
	\end{tabular}
	\caption{Non-informative priors for claim frequency and claim severity.}\label{tab:non_inform}
\end{table}
\end{comment}
\begin{table}[H]
	\centering
	\small
	\begin{tabular}{l|cccc}
		\hline\hline
		Model&\MyHead{1.8cm}{Model\\ Para.}&Prior&\MyHead{1.8cm}{Hyper \\Para.}& Posterior Para.\\
		\hline
		\multirow{2}{*}{Poisson}&\multirow{2}{*}{$\mu$} &\multirow{2}{*}{Gamma}&$\alpha_0$&$\alpha_0+n$\\
		&&&$\beta_0$&$\beta_0/(\beta_0 A+1)$\\
		\hline
		\multirow{2}{*}{Gamma}&\multirow{2}{*}{$(1/\beta)|\alpha$}&\multirow{2}{*}{Gamma}&$\alpha_0$&$\alpha_0+\alpha n$\\
		&&&$\beta_0$&$[\sum\limits_{i=1}^{n}y_i+1/\beta_0]^{-1}$\\
		\hline
		\multirow{5}{*}{Lognormal}&\multirow{3}{*}{$1/\sigma^2$}&\multirow{3}{*}{Gamma}&$\alpha_0$ &$\alpha_0+n/2$\\
		&&&\multirow{2}{*}{$\beta_0$}&$[1/\beta_0+(n-1)/2\Var(\ln y)+$\\
		&&&&$\qquad\qquad \kappa_0/(2(n+\kappa_0))(\E(\ln y)-\xi_0)^2]^{-1}$\\
		\cline{2-5}
		&\multirow{2}{*}{$\xi|\sigma^2$}&\multirow{2}{*}{Normal}&$\xi_0$&$n\E(\ln y)/(\kappa_0+n)+\kappa_0\xi_0/(\kappa_0+n)$\\
		&&&$\kappa_0$&$\sigma^2/(\kappa_0+n)$\\
		\hline
	\multirow{2}{*}{Pareto}&\multirow{2}{*}{$\alpha|\beta$}&\multirow{2}{*}{Gamma}&$\alpha_0$&$\alpha_0+n$\\
	&&&$\beta_0$&$[\sum\limits_{i=1}^n\log(1+y_i/\beta)+1/\beta_0]^{-1}$\\
		\hline\hline
	\end{tabular}
\caption{Conjugate priors for the claim frequency and the claim severity distribution.}\label{tab:conju}
\end{table}

%Table \ref{subsec:5.1.1} lists the prior distributions for each parameter in the Poisson claim frequency distribution and Gamma, Lognormal and Pareto claim severity distributions. In the frequentist method, the intensity parameter $\mu$ in the Poisson distribution is treated as a known quantity. In order to reflect this strong belief, we therefore assume a Gamma conjugate prior for $\mu$ in the Bayesian setting. The claim severities, on the other hand, are more difficult to model, so both the non-informative and informative priors are utilised. Conjugate priors can be easily constructed for Lognormal distribution, but there is no conjugate prior for the shape parameter $\alpha$ in both the Gamma and the Pareto distribution. For simplicity, Gamma distribution are used as the prior distribution for the shape parameter $\alpha$. Also, it is worth noting that Gamma distribution is the conjugate prior for $1/\sigma^2$ in the Lognormal distribution, therefore $\sigma^2$ instead of $\sigma$ is used for computational convenience. The Jeffreys prior is used except for the Lognormal distribution, where the so-called reference prior is assumed. It is just a special case of the conjugate prior, with the hyper parameters $\kappa_0=0, \nu_0=-1$ and $\sigma_0=0$. The simulation process in \eqref{equ:5.6} is carried out through MCMC, either by the Gibbs sampler or Metropolis-Hastings method. 
%and it allows the use of Gibbs sampling instead of Metropolis-Hastings(MH).

\subsection{Numerical results}\label{subsec:5.3}
Table \ref{tab:hypers} presents values of the true parameters $(\mu_0,\B\zeta_0)$ and the hyper parameters in the informative priors. We investigate how the results are influenced by varying the historical portfolio size $J_h$ from $10^5$ to $10^3$, which corresponds to have an expected sample size $n$ from $5000$ to $50$. The simulation results for both the non-informative and informative priors are shown in Table \ref{fig:bayes_conju} and \ref{fig:bayes_jeffrey}, with the reinsurance  priced according to the expected premium principle and the VaR as the risk measure. 

The optimal ratio $C(\bf a; \mu_0, \B\zeta_0)$ under the true parameters is presented in the first column, below each candidate model. As expected, the error increases when the sample size decreases regardless of the prior choice, but the rate of degradation seems unclear from the two tables.
In general, the informative prior gives smaller $\E[D(\mu_0,\B{\zeta}_0)]$ and $\sd[D(\mu_0,\B{\zeta}_0)]$ compared to the non-informative ones, especially when there are few observations. When sample size is large, the posterior distribution is robust to prior assumptions. With a limited amount of data, the value of the error can be still accepted if there is strong knowledge of the prior.
Compared with the results of the bootstrap method in Table \ref{tab:boot_expt}, the discrepancy between these two methods is minor when data are sufficient. However, the errors evaluated using the Bayesian method with an informative prior are relatively smaller than the bootstrap method when the sample size is small. In a word, with strong prior belief about the parameters, the Bayesian method might be preferred to the frequentist method, in particular with limited historical data. The results for the mixed Esscher premium principle are quite similar and not included here. 
%As elaborated in Section \ref{sec:uni_error}, the claim severity distributions do not seem very important as long as they are calibrated to have  a common mean and standard deviation, since the estimated $\hat{C}_I$ is essentially identical, but it is not true for the Bayesian method. These tables demonstrate that 
%the errors assessed by the Bayesian method are substantial for heavy-tailed Lognormal and Pareto distribution, especially for smaller sample size.  

\begin{table}[H]
	\centering
	\small
	\begin{tabular}{l|lll}
		\hline\hline
		Model&Para.&$(\mu_0,\B\zeta_0)$& Hyper Para.\\
		\hline
		Poisson&$\mu$&$0.05$&$(\alpha_0,\beta_0)=(0.25,0.2)$\\
		\hline
		\multirow{2}{*}{Gamma}&$\alpha$&0.44&$(\alpha_0,\beta_0)=(10,0.1)$\\
		&$\beta$&22.50&$(\alpha_0,\beta_0)=(1,0.1)$\\
		\hline
		\multirow{2}{*}{Lognormal}&$\xi$&1.71&$(\xi_0,\kappa_0)=(2,100)$\\
		&$\sigma$&1.09&$(\alpha_0,\beta_0)=(8,0.1)$\\
		\hline
		\multirow{2}{*}{Pareto}&$\alpha$&3.60&$(\alpha_0,\beta_0)=(40,0.1)$\\
		&$\beta$&26.00&$(\alpha_0,\beta_0)=(3000,0.01)$\\
		\hline\hline
	\end{tabular}
	\caption{The values of true and hyper parameters in claim frequency and claim severity distributions}\label{tab:hypers}
\end{table}

\begin{table}[H]
	\centering
	\small
	\begin{tabular}{llccccc}
		\hline
		\hline
		Model & $J_h$& $\E(a_1^*)$ &$\E(a_2^*)$ & $\E(C({\hat{\bf{a}}}_B, \mu_0,\B{\zeta}_0))$ &$\E[D(\mu_0,\B{\zeta}_0)]$&$\sd[D(\mu_0,\B{\zeta}_0)]$\\
		\hline
		\multirow{3}{*}{\parbox{2cm}{Gamma\\$(12.46)$}}&  $10^5$&  523.2&838.5&12.63&  0.173&   0.2405\\
		&$10^4$& 529.0&  867.7& 12.75& 0.294 &  0.5087 \\
		&$10^3$& 530.8&975.7&13.62&1.158 &2.1511\\
		\hline
		\multirow{3}{*}{\parbox{2cm}{Lognormal\\$(12.40)$}}& $10^5$& 516.9& 868.6& 12.55&   0.151&0.2344 \\
		&$10^4$& 510.9&  864.7& 12.92&  0.523& 0.6417  \\
		&$10^3$& 491.9&  851.9& 13.25 & 0.847&1.3896 \\
		\hline
		\multirow{3}{*}{\parbox{2cm}{Pareto\\$(12.37)$}}&
		$10^5$& 518.7& 865.8& 12.56&0.188& 0.2642\\
		&$10^4$& 520.2& 892.1&13.00&0.626 &0.8994\\
		&$10^3$& 514.6&997.1&13.93& 1.556&2.4950 \\
		\hline\hline
	\end{tabular}
	\caption{Bayesian errors with the informative priors for $(\mu,\B{\zeta})$ under VaR when premium is calculated based on expected premium principle.}\label{fig:bayes_conju}
\end{table}

\begin{table}[H]
	\centering
	\small
	\begin{tabular}{llccccc}
		\hline
		\hline
		Model & $J_h$& $\E(a_1^*)$ &$\E(a_2^*)$ & $\E(C({\hat{\bf{a}}}_B, \mu_0,\B{\zeta}_0))$ &$\E[D(\mu_0,\B{\zeta}_0)]$&$\sd[D(\mu_0,\B{\zeta}_0)]$\\
		\hline
		\multirow{3}{*}{\parbox{2cm}{Gamma\\$(12.46)$}}&
		$10^5$& 522.9&  827.5&12.74 &0.281&  0.2589\\
		&$10^4$& 525.6& 850.0&12.94&0.486 & 0.8025\\
		&$10^3$& 554.3& 1095.6&14.32 &1.860 &3.3671\\
		\hline
		\multirow{3}{*}{\parbox{2cm}{Lognormal\\$(12.40)$}}& $10^5$& 517.2&  871.0& 12.61&  0.209& 0.3041\\
		&$10^4$& 520.2&  895.4& 12.89&  0.694 &  0.9066\\
		&$10^3$& 546.4&1160.8&14.25&1.851&3.4457\\
		\hline
		\multirow{3}{*}{\parbox{2cm}{Pareto\\$(12.37)$}}& $10^5$& 516.6& 861.3& 12.63& 0.257& 0.3491 \\
		&$10^4$& 508.1& 862.9&13.15&0.777 &1.0834\\
		&$10^3$&519.7& 1005.5&15.00&2.625&3.4847\\
		\hline\hline
	\end{tabular}
	\caption{Bayesian errors with the non-informative priors for $(\mu,\B{\zeta})$ under VaR when premium is calculated based on expected premium principle.}\label{fig:bayes_jeffrey}
\end{table}

\section{Concluding discussion}\label{sec:con}
This paper have discussed the reinsurance optimization problem by examining how much the solution is affected by errors in the parameters and models. Since the single layer contracts are often optimal or close to optimal in many situations, it is of great relevance to examine how estimation errors degrade the single layer solutions. More specifically, the problem is formulated under a more general reinsurance pricing function and an industrially plausible criterion, which is the ratio of VaR(or CVaR) against the expected gain of the cedent. Then the degradation of the single layer contracts is investigated through both asymptotics and numerical studies. It is shown that the rate of degradation is often $O(1/n)$ as the the sample size $n$ of historical observations becomes infinite, but criteria based on VaR are exceptions that may
achieve only $O(1/\sqrt{n})$. This result is verified in the numerical study. We also point out that the choice of the claim severity distribution does not seem important on determining the optimal ratio, given that the distributions are calibrated to have the same mean and standard deviation. In this case, the simple Gaussian distribution for the risk might be a reasonable alternative.
The Bayesian approach offers a different way of estimating and evaluating parameter errors. The numerical results are similar to what we have in the frequentist method when there are sufficient data. While with a limited amount of data, the errors from the Bayesian approach are relatively smaller, but they depend on how much prior information we have for the parameters. 
One question has not been tackled is how the model uncertainty or error influences the optimal solutions. It is worth investigating the impact when the true family of claim severity distribution deviates from the assumed one. Whether or not the Gaussian distribution is a sensible approximation for the total loss in this optimization problem can be examined through a similar manner. We leave this for future research.

\bibliography{optre3_new.bib}

\appendix
%\section*{Appendix}
\section*{Appendices}
\numberwithin{equation}{subsection}
\renewcommand{\thesubsection}{\Alph{subsection}}
\subsection{Proofs of  degradation asymptotics}\label{appen:a}
\subsubsection{Proof of Proposition \ref{prop3.1}.}\label{proofA1}
The  argument is a standard one using Taylor expansions.
Indeed, from~(\ref{e33})
\begin{displaymath}
D(\B\theta)=\sum_{i=1}^{n_a}
\frac{\partial C(\bf{a},\B\theta)}{\partial a_i}(\hat{a}_i-a_i)
+\frac{1}{2}\sum_{i=1}^{n_a}\sum_{j=1}^{n_a}\frac{\partial^2 C(\bf{a},\B\theta)}{\partial a_i\partial a_j}
(\hat{a}_i-a_i)(\hat{a}_j-a_j)+{\cal E}_1
\end{displaymath}
with ${\cal E}_1$ a remainder term. 
The linear term vanishes since the partial derivatives are zero 
at the minimum. Hence, with the matrix $C^{aa}$ introduced in Section \ref{subsec:asy_smooth},
\begin{equation}
D(\B\theta)=\frac{1}{2}(\hat{{\bf a}}-{\bf a})^TC^{aa}(\hat{{\bf a}}-{\bf a})
+{\cal E}_1,
\label{a1}
\end{equation}
where we must replace $\hat{{\bf a}}-{\bf a}$ with its relationship to
$\hat{\B\theta}-\B\theta$. Note that
\begin{displaymath}
\frac{\partial C({\bf a},\hat{\B\theta})}{\partial a_i}
=\frac{\partial C({\bf a},\B\theta)}{\partial a_i}
+\sum_{j=1}^{n_a}\frac{\partial^2 C({\bf a},\B\theta)}{\partial a_i\partial a_j}(\hat{a}_j-a_j)
+\sum_{k=1}^{n_\theta}\frac{\partial^2 C({\bf a},\B\theta)}{\partial a_i\partial \B\theta_k}
(\hat{\B\theta}_k-\B\theta_k)+{\cal E}_{i2}
\end{displaymath}
with ${\cal E}_{i2}$  another remainder. 
Both first order derivatives are zero so that on matrix form
this may be rewritten
\begin{displaymath}
C^{aa}(\hat{{\bf a}}-{\bf a})+C^{a\theta}(\hat{\B\theta}-\B\theta)
+{\cal E}_2=0
\end{displaymath}
with ${\cal E}_2=({\cal E}_{12},\dots,{\cal E}_{n_a2})^T$ and
where $C^{a\theta}$ was defined above. The matrix $C^{aa}$ is the second 
order derivatives  at a minimum 
and is therefore positive definite and can be inverted.
This yields
\begin{displaymath}
\hat{{\bf a}}-{\bf a}=-(C^{aa})^{-1}C^{a\theta}(\hat{\B\theta}-\B\theta)
-(C^{aa})^{-1}{\cal E}_2,
\end{displaymath}
and when this is inserted for $\hat{{\bf a}}-{\bf a}$ in the expression for
$D(\B\theta)$, it follows that
\begin{displaymath}
D(\B\theta)=(\hat{\B\theta}-\B\theta)^TQ(\hat{\B\theta}-\B\theta)
+{\cal E}_1+\frac{1}{2}{\cal E}_2^T(C^{aa})^{-1}{\cal E}_2
\end{displaymath}
with $Q$ as in~(\ref{e36}) right.
It follows that
the  asymptotic distribution of $D(\B\theta)$
is that of a quadratic form under normal variables, as stated above
if the remainder terms vanish. We need to argue that
$n{\cal E}_1\rightarrow 0$ and $\sqrt{n}{\cal E}_2\rightarrow 0$ 
as $n\rightarrow \infty$, and both limits
are consequences of second order derivatives being uniformly bounded
.
\subsubsection{Proof of Proposition \ref{prop3.2}.}\label{proofA2}
Let $\rho_{R_i}$ in the risk over expected gain criterion~(\ref{e210})
be Value at Risk which is now denoted
 $R({\bf a},\B\theta)$ under the one-layer contract $I=I_{\bf a}$ so that
$C({\bf a},{\B\theta})=R({\bf a},\B\theta)/G({\bf a},\B\theta)$.
It is convenient to proceed in terms of
\begin{equation}
C_0({\bf a},\B\theta)=
\frac{R({\bf a},\B\theta)}
{G_0({\bf a},\B\theta)}
\qquad\mbox{where}\qquad
G_0({\bf a},\B\theta)=G({\bf a},\B\theta)
	+\beta R({\bf a},\B\theta)
\label{a2}
\end{equation}
with degradation
\begin{equation}
D_0(\B\theta)=
C_0(\hat{\bf a},\B\theta)-
C_0({\bf a},\B\theta).
\label{a3}
\end{equation}
Note that
\begin{displaymath}
C({\bf a},\B\theta)=
\frac{C_0({\bf a},\B\theta)}
{1-\beta C_0({\bf a},\B\theta)},
\end{displaymath}
and $C_0({\bf a},\B\theta)$ and $C({\bf a},\B\theta)$ have minimum
at the same $\bf a$ while the original degradation 
$D(\B\theta)=
C(\hat{\bf a},\B\theta)-
C({\bf a},\B\theta)$ has a simple asymptotic relationship to
$D_0(\B\theta)$. Indeed,
\begin{displaymath}
D(\B\theta)=\frac{C_0(\hat{\bf a},\B\theta)}
{1-\beta C_0(\hat{\bf a},\B\theta)}-\frac{C_0({\bf a},\B\theta)}
{1-\beta C_0({\bf a},\B\theta)}
=\frac{ C_0(\hat{\bf a},\B\theta)- C_0({\bf a},\B\theta)}
{\{1-\beta C_0({\bf a},\B\theta)\}^2}+o_p(1/\sqrt{n})
\end{displaymath}
after a Taylor argument around $C_0({\bf a},\B\theta)$. The error term
$o_p(1/\sqrt{n})$ comes from the discrepancy $\hat{\bf a}-{\bf a}$
being of order $1/\sqrt{n}$. Hence after inserting for
$C_0(\hat{\bf a},\B\theta)$ and $C_0({\bf a},\B\theta)$
it follows that
\begin{equation}
D(\B\theta)=\{1+\beta C({\bf a},\B\theta)\}^2
D_0(\B\theta)+o_p(1/\sqrt{n}),
\label{a4}
\end{equation}
and the asymptotic distribution of
$D(\B\theta)$ is inherited from 
that of $D_0(\B\theta)$.

Value at risk for the insurer under $\hat{\bf a}$ and $\bf a$ are
\begin{displaymath}
R({\bf a},\B\theta)=a_1
\quad\mbox{and}\quad
R(\hat{\bf a},\B\theta)=
(x_\epsilon(\B\theta)-\hat{a}_2)_++\hat{a}_1
=(a_2-\hat{a}_2)_++\hat{a}_1,
\end{displaymath}
so that
\begin{displaymath}
D_0(\B\theta)=
\frac{(a_2-\hat{a}_2)_+}
{G_0(\hat{\bf a},\B\theta)}
+\left(\frac{\hat{a}_1}
{G_0(\hat{\bf a},\B\theta)}
-\frac{a_1}
{G_0({\bf a},\B\theta)}\right),
\end{displaymath}
and the second term has to be linearized. When
Taylor's formula is invoked 
around $(a_1,a_2)$, the linear term in $\hat{a}_1-a_1$ vanishes
since the partial derivative is $0$ at the optimum $a_1$, but the 
second partial derivative must be calculated. Recall that
\begin{displaymath}
G_0({\bf a},\B\theta)=\gamma\pi-\int_{a_1}^{a_2}K\{F(x)\}dx,
\end{displaymath}
from which it follows that
\begin{displaymath}
\frac{\partial}{\partial a_2}\left(\frac{a_1}
{G_0({\bf a},\B\theta)}\right)=-\frac{a_1}
{G_0({\bf a},\B\theta)^2}\frac{\partial G_0({\bf a},\B\theta)}
{\partial a_2}
=\frac{a_1K\{F(a_2,\B\theta)\}}
{G_0({\bf a},\B\theta)^2}
=\frac{a_1K(1-\epsilon)}
{G_0({\bf a},\B\theta)^2},
\end{displaymath}
since $F(a_2,\theta)=1-\epsilon$. Hence
\begin{displaymath}
D_0(\B\theta)=\frac{(a_2-\hat{a}_2)_+}
{G_0({\bf a},\B\theta)}+\frac{a_1K(1-\epsilon)}
{G_0({\bf a},\B\theta)^2}(\hat{a}_2-a_2)+o_p(1/\sqrt{n}).
\end{displaymath}
Note that
\begin{displaymath}
\frac{1}{G_0({\bf a},\B\theta)}
=\frac{1}{G({\bf a},\B\theta)+\beta R({\bf a},\B\theta)}
=\frac{C({\bf a},\B\theta)/a_1}{1+\beta C({\bf a},\B\theta)},
\end{displaymath}
and when this with the  expression for $D_0(\B\theta)$ are 
inserted into~(\ref{a4})
some straightforward calculations yield
\begin{equation}
D(\B\theta)=
h_1({\bf a},\B\theta)(a_2-\hat{a}_2)_+
+h_2({\bf a},\B\theta)(\hat{a}_2-a_2)+o_p(1/\sqrt{n}),
\label{a5}
\end{equation}
where
\begin{displaymath}
h_1({\bf a},\B\theta)=\{1+\beta C({\bf a},\B\theta)\} C({\bf a},\B\theta)/a_1
\quad\mbox{and}\quad
h_2({\bf a},\B\theta)=C({\bf a},\B\theta)^2K(1-\epsilon)/a_1
\end{displaymath}
are the coefficients in~(\ref{e38a}).

The asymptotic properties of $D(\B\theta)$ follows from the 
representation~(\ref{a5}).
Recall that $\hat{a}_2-a_2=x_\epsilon(\hat{\B\theta})-x_\epsilon(\B\theta)$
for which 
a standard argument shows that
\begin{displaymath}
\sqrt{n}\{
x_\epsilon(\hat{\B\theta})-x_\epsilon(\B\theta)\}
\stackrel{d}{\longrightarrow} {\bf g}^T{\bf N}=V\sqrt{{\bf g}^T\Sigma{\bf g}^T}
\end{displaymath}
with $\bf g$ the gradient vector~(\ref{e38}), $\bf N$ the normal 
vector in~(\ref{e34}) and $V$ as in~(\ref{e38b}).
Since $\bf N$ has mean zero expectations and covariance matrix
$\Sigma$, it follows that  $V$ is the standard normal.
Slutsky's theorem applied to~(\ref{a5}) 
with $\hat{a}_2-a_2$ replaced by 
$x_\epsilon(\hat{\B\theta})-x_\epsilon(\B\theta)$
now yields the limit 
\begin{displaymath}
\sqrt{n}D(\B\theta)\stackrel{d}{\longrightarrow}\{
h_1({\bf a},\B\theta)(-V)_+
+h_2({\bf a},\B\theta)V\}\sqrt{{\bf g}^T\Sigma{\bf g}},
\end{displaymath}
which is~(\ref{e39}) in 
Proposition \ref{prop3.2}. 

To verify the expressions~(\ref{e310})
and~(\ref{e310a}) for the mean and
variance let
 $\varphi(v)=(2\pi)^{-1/2}e^{-v^2/2}$ and note
that
\begin{displaymath}
E(-V)_+=\int_{-\infty}^0(-v)\varphi(v)=1/\sqrt{2\pi}
\quad\mbox{whereas}\quad E(V)=0,
\end{displaymath}
and~(\ref{e310}) follows. For the variance 
\begin{displaymath}
E\{h_1({\bf a},\B\theta)(-V)_+
+h_2({\bf a},\B\theta)V\}^2=E\{h_1({\bf a},\B\theta)^2(-V)_+^2
+2h_1({\bf a},\B\theta)h_2({\bf a},\B\theta)(-V)_+V
+h_2({\bf a},\B\theta)V^2\}
\end{displaymath}
\vspace*{-0.5cm}
\begin{displaymath}
\hspace*{4.80cm}=h_1({\bf a},\B\theta)^2\frac{1}{2}
+2h_1({\bf a},\B\theta)h_2({\bf a},\B\theta)\frac{1}{2}+
h_2({\bf a},\B\theta)^2,
\end{displaymath} 
and hence
\begin{displaymath}
\Var\{h_1({\bf a},\B\theta)(-V)_+
+h_2({\bf a},\B\theta)V\}
=h_1({\bf a},\B\theta)^2/2
+h_1({\bf a},\B\theta)h_2({\bf a},\B\theta)+
h_2({\bf a},\B\theta)^2-E\{(-V)_+\}^2
\end{displaymath}
\vspace*{-0.5cm}
\begin{displaymath}
\hspace*{4.80cm}=\frac{1}{2}(1-1/\pi)h_1({\bf a},\B\theta)^2
+h_1({\bf a},\B\theta)h_2({\bf a},\B\theta)+
h_2({\bf a},\B\theta)^2,
\end{displaymath}
which yields~(\ref{e310a}).
\subsection{Simulation Algorithms}\label{appen:b}
\begin{algorithm}[H]
	\small
	\caption{The total loss and its quantile simulation\label{alg:total_loss}}
	Input: $m$, $\epsilon$, $J$, $\mu$, $T$, $f_{Y}(\cdot;\B{\zeta})$
	\begin{algorithmic}[1]
		\For{$i = 1,\ldots,m$}
		\State $X_{i}^{*} = 0$
		\State Draw $N^{*} \sim \textrm{Poisson}(J\mu T)$
		\For{$j = 1,\ldots,N^{*}$}
		\State Draw $Y^{*}$ from $f_{Y}(y;\B{\zeta})$
		\State $X_{i}^{*} = X_{i}^{*}+Y^{*}$
		\EndFor
		\EndFor
		\State Sort as $X_{(1)}^{*}\leq\ldots\leq X_{(m)}^{*}$
		\State  $q_{\epsilon}=X_{((1-\epsilon)m)}^{*}$
		\State \Return $q_\epsilon$
	\end{algorithmic}
\end{algorithm}
\vspace*{-1.5em}
\begin{algorithm}[H]
	\small
	\caption{The nested bootstrap algorithm\label{alg:boots_crit}}
	Input: $m, J, T, \hat\mu, \hat{\B{\zeta}}$, $f_{Y}(\cdot;\B{\zeta}), C(\B{{\bf a}},\B{\theta})$
	\begin{algorithmic}[1]
		%\For{$i = 1,\ldots,m_b$}
		\State Draw $\hat{N}^{*} \sim \textrm{Poisson}(J\hat\mu T)$, $\hat{Y}^*_1,\ldots,\hat{Y}^*_n \sim f_Y(y,\hat{\B{\zeta}})$
		\State $\hat\mu^*\xleftarrow{MLE} \hat{N}^{*}/JT$, $\hat{\B{\zeta}}^* \xleftarrow{MLE}\hat{Y}^*_1,\ldots,\hat{Y}^*_n$
		\For{$i = 1,\ldots,m$}
		\State $X^{*\star}_i=0$
		\State Draw $\hat{N}^{*\star} \sim \textrm{Poisson}(J\hat\mu^* T)$
		\For{$k = 1,\ldots,\hat{N}^{*\star}$}
		\State Draw $\hat{Y}^{*\star}$ from $f_{Y}(y;\hat{\B{\zeta}}^*)$
		\State $X_{i}^{*\star} = X_{i}^{*\star}+\hat{Y}^{*\star}$
		%\State$\hat{\bf{a}}^*=\underset{{\bf{b}}}{\argmin}\;   C({\bf{b}},\hat{\B{\theta}}_i^*)$
		\EndFor
		\EndFor
		%\EndFor
		\State \Return $\hat{\bf{a}}^*=\underset{{\bf{b}}}{\argmin}\;   C({\bf{b}},\hat{\B{\theta}}_i^*)$
	\end{algorithmic}
\end{algorithm}
\vspace*{-1.5em}
\begin{algorithm}[H]
	\small
	\caption{Bayesian method for degradation evaluation\label{alg:bayes_crit}}
	Input:  $ m, J_h, J, T, \mu_0, \zeta_0, f_{Y}(\cdot;\B{\zeta}), p(\mu|n), p(\bm\zeta|\B y), C({\bf a}, \mu, \B\zeta)$
	\begin{algorithmic}[1]
		%\For{$i = 1,\ldots,m_b$}
		\State Draw $n \sim \mathrm{Poisson}(J_h\mu_0 T), \;\bm y \sim f_{Y}(y;\B{\zeta}_0)$
		\For{$i = 1,\ldots,m$}
		\State $X^{\star}_i=0$
		\State Draw $\mu^\star \sim p(\mu|n), \;\bm\zeta^\star \sim p(\bm\zeta|\B y)$
		\State Draw $N^{\star} \sim \textrm{Poisson}(J\mu^\star T)$
		\For{$k = 1,\ldots,N^{\star}$}
		%\For{$k = 1,\ldots,N^{*}$}
		\State Draw $Y^{\star}$ from $f_{Y}(y; \B\zeta^\star)$
		\State $X_{i}^{\star} = X_{i}^{\star}+Y^{\star}$,
		\EndFor
		\EndFor 
		%\State $\hat{\bf{a}}_i^\star=\underset{{\bf b}}{\argmin}\;    C({\bf b},\mu^\star, \B \zeta^\star)$
		%\EndFor
		\State \Return $\hat{\bf{a}}_B^\star=\underset{{\bf b}}{\argmin}\;    C({\bf b},\mu^\star, \B \zeta^\star)$
	\end{algorithmic}
\end{algorithm}
\end{document}